\documentclass[12pt]{article}
\usepackage{graphicx}
\begin{document} 

\bigskip
\centerline
{\bf \Large
Modelling linguistic taxonomic dynamics}

\bigskip
\noindent
S{\o}ren Wichmann$^{1,2}$, Dietrich Stauffer$^3$, F. Welington S. Lima$^4$, and Christian Schulze$^3$

\bigskip

\noindent
{\small$^1$ Department of Linguistics, Max Planck Institute for Evolutionary Anthropology, Deutscher Platz 6, D-04103 Leipzig, Germany

\medskip
\noindent
$^2$ Languages and Cultures of Indian America (TCIA), P.O. Box 9515, 2300 RA  Leiden, The Netherlands

\medskip
\noindent
$^3$ Institute for Theoretical Physics, Cologne University, D-50923 K\"oln, Euroland

\medskip
\noindent
$^4$ Departamento de F\'{\i}sica, Universidade Federal do Piau\'{\i}, 57072-970 Teresina - PI, Brazil}

\bigskip
Abstract

{\small This paper presents the results of the application of a bit-string model of languages (Schulze and Stauffer 2005) to problems of taxonomic patterns. The questions addressed include the following: (1) Which parameters are minimally needed for the development of a taxonomic dynamics leading to the type of distribution of language family sizes currently attested (as measured in the number of languages per family), which appears to be a power-law? (2) How may such a model be coupled with one of the dynamics of speaker populations leading to the type of language size seen today, which appears to follow a log-normal distribution?}

\section{Introduction}

With few exceptions, such as Nettle (1999a,b), linguists have been little concerned with quantitative modeling and simulation, possibly due to the myriad of qualitative phenomena that scholars must analyze. An immense amount of structural differences exist not only from one language to the next, but also among different kinds of sociolinguistic situations. More recently, however, scholars belonging to the entirely different discipline of physics have taken an interest in simulating the aspect of historical sociolinguistics which concerns the competition among languages and have looked at how such competition may lead to various patterns of growth or extermination (see Schulze and Stauffer 2006 for a review). This interest among physicists for modeling language competition was triggered by Abrams and Strogatz (2003), who use differential equations to describe the vanishing of one language due to the dominance of another. Since then, a series of articles have appeared. For instance, Patriarca and Lepp\"anen (2004) applied the Abrams-Strogatz model to a geographical situation where two languages, X and Y, may dominate in each their region, resulting in the survival of both, rather as in the original model where only one language will survive. Oliveira et al. (2006a,b) have looked at models in which speakers of small languages will tend to switch to geographically more widespread ones, to account for the fact that real geographical areas tend not to show equally-sized languages. Along the lines of the broader cultural model of Axelrod (1997), Te\c sileanu and Meyer-Ortmanns (2006) looked at the consequences of the possibility that a greater similarity among languages might further language shift. We build on some of this work, but so far the present paper is to our knowledge the first in this recent tradition to address the issue of taxonomic dynamics.

\section{Why simulate}

Human languages have existed for at least about $10^4$ years, possible much longer. Only a few percent of this development is to some extent documented through writing, while another few percent may be inferred by comparative linguistic methods. Thus, we have no clues to aspects of the development of languages for 80\% or more of their history other than what we might infer from abstract extrapolation or from simulations. Like the distant past, the future is also empirically impenetrable. Two aspects of simulations are important. First, we may hope to identify a minimal number of parameters that account for the present state of affairs seen as a result of a long development. Secondly, we may adjust these parameters to test the predictions that different models provide. It should be stressed that simulations are not necessarily suited to prove any particular model or to make predictions about what is in store for the languages of today; they can only represent tests of different models. Nor can the  parameters identified be translated into direct explanatory factors for actual distributions. For instance, a simulation of language competition might restrict its parameters to, say, the relative size of languages and it might stipulate some simple mode of interaction, such as the tendency for speakers of smaller languages to shift to contiguous larger ones. Such a model might lead to a plausible picture of language distributions, perhaps even one resembling the current state of affairs. But this does not mean that this distribution is explained by language sizes and competition alone. The growth of a given language relates to socioeconomic, historical, geographical, ecological and many other circumstances. Since a primary aim of simulation is to reduce the set of parameters, it cannot and should not, however, take into account all relevant factors, but must remain an abstraction.

\section{The aim of the investigation}

Wichmann (2005) made some simple observations about the present-day quantitative distribution of language family sizes, as measured in numbers of languages per family, and about the distribution of language sizes, as measured in numbers of speakers per language (data drawn from \textit{Ethnologue}). It was found that language family sizes approximate a so-called `power-law', that is, a distribution described by the equation $y = ax^b$, which corresponds to a straight line on a log-log plot. Such distributions are frequent in both nature and the social world (cf. Newman 2005 for an excellent overview). The slope of the curve on the rank-by-size plot is described by the exponent $b$, which was found to be $-1.905$. (For a histogram of the number $n(S)$ of languages versus their size S this corresponds to another power-law with exponent $-1-1/1.905$, and if this histogram sums the raw numbers into bins whose size is proportional to the language size, then the exponent is $-1/1.905$.)

When testing for the distribution of language sizes, however, no power-law emerged. The absence of a power-law distribution also comes out of studies by Novotny and Drozd (2000) and Sutherland (2003). (Gomes et al. 1999: 493) had earlier plotted the same data on a graph showing the cumulative size distribution, $n( > S)$, corresponding to the number of languages with a size greater than S. Cutting the curve up into different regions and describing each by a separate equation they then made the problematical claim of the existence of a ``composite power-law''). The present paper takes up the challenge of Wichmann (2005: 139) to test, using computer simulations, what the expected past and future distributions of language family sizes and language sizes might look like. The question was raised whether the present distribution of language sizes might be characteristic of a stage of disequilibrium while the expected equilibrium might correspond to a power-law. Stauffer et al. (2006) supported the hypothesis of a disequilibrium. In the present paper we also report on language families.

\section{The bit-string model}

The model used is one eminently suited to computation. It is a variant of that of Schulze and Stauffer (2005), which operates with bit-strings of length L, where each bit has two values and where the total set of possible dialects has $2^L$ members. (A precursor to this kind of modelling is Wang and Minett 2005, which used strings of integers to simulate branching by the mutation and transfer of numbers.) 

Each bit may be interpreted as the presence or absence of some characteristic grammatical feature. Under this interpretation we might imagine that a number of diagnostic features were identified, the presence or absence of each of which would be sufficient to distinguish among the grammars of the world's languages. This number corresponds to the length of the bit-string.

An alternative model of language competition, also allowing for thousands of different languages, is that of de Oliveira et al (2006a,b). There, however, languages are characterized merely by consecutive numbers 1, 2, 3, ..., which is not suitable for simulating different taxonomic levels. In this model, language families would have to be determined by the history of language dynamics and their genealogical tree (Schulze and Stauffer 2006), and testing this approach is outside the scope of the present work. The other recent models of language competition to our knowledge allow only a relatively small number of languages and are, for this reason, also less suitable for taxonomy.

We test two different variants of the model. In one, which we might call the ``hierarchical'' variant, the bit-string is divided into subsections corresponding to different taxonomic levels. Two languages are defined as belonging to the same family if their ``family'' parts of the bit-strings agree. In the other, ``flat'' variant, there is no such partitioning of the string. Instead, taxonomic levels are achieved by defined a certain threshold $k$ of differences among languages. Differences are measured by comparing two strings and noting the number of positions for which the two strings differ. If the difference is greater than $k$, the two languages are said to belong to different taxa. In both versions of the model we only operate with two taxonomic levels, but both could be extended to include more levels. In the following, each variant is described in more detail.

\subsection{The hierarchical variant}

This model achieves two taxonomic levels by partitioning the bit-string. The two levels may be conceptualized as corresponding to language families and languages within one family, respectively, but need not be translated exactly into these concepts (which are themselves not very well defined). The languages of individuals may be classified by comparing the bit-strings representing each individual. In the following we illustrate how the model works if we use a bit-string of length 64. People speaking the same language have to agree in all bits. In our implementation we have chosen to stipulate that people speaking languages belonging to the same family have to agree in the leftmost 19 bits. For example, 

\medskip
\noindent
01101010011110101010-11101010010101101010010111010101010011100001

\noindent
01101010011110101010-11101010010101101010010111010101010011101001

\medskip
\noindent
are two different (even if potentially closely related) languages, while

\medskip
\noindent
00101001101010100011-10100101011011010101110101010110101010100111

\noindent
00101001101010100011-01011010001011000110010101110000110101001100

\medskip
\noindent
are two different languages belonging to the same family and 

\medskip
\noindent
10100110100101011010-10010110101010101111010101101001110001010001

\noindent
01011011010101101010-10100011110101011010101000110101010011101101

\medskip
\noindent
are two different languages belong to two different families. In these examples the dash ``-'' just indicates the boundary between the two segments of the string, analogously to the convention for phone numbers, which are structured much like our bit-string model.

The choice of lengths of the whole string and its parts is of course arbitrary, and need not be 19 + 45 = 64. Nevertheless, various considerations led to single out certain lengths as more suitable than other. First, the model is computationally most effective for bit-string lengths which are powers of 2. Second, a shorter string is to be preferred to a longer one, all else being equal---again for computational reasons. Third, the string should not be so short that the sheer length imposes artificial constraints on the  results. In earlier simulations the effect of different lengths ($L = 8$, 16, 32, and 64) were tested. Since it was found that the results were qualitatively similar for $L = 16$, 32, and 64, all values of $L$ higher than or equal to 16 would be equally suitable. Adding the criterion of minimal computation cost would single out $L = 16$ as preferable. However, we found that for this length a maximum number of languages was reached before a meaningfully interpretable distribution was found. (Unlike the 32 and 64 bit-string models and the real-life present-day distribution, see section 5 below, this did not lead to a power-law distribution, since power-laws require the absence of upper bounds. At a point where either all possible languages or all possible families are filled, the power-law distribution breaks down.) Instead, we have chosen a string with the larger length of 32 bits, of which the leading 10 bits define families and the remaining 22 define languages, yielding $2^{10} = 1024$ possible `families' and $2^{22} = 4,194,304$ posssible `languages'). Using an ample $L$ also ensures that accidental `back mutation', i.e. the phenomenon whereby, by chance, an identical bit-string occurs after some mutations--something which would not happen in real life--will occur so exceedingly rarely that its effects are completely negligible. (Even for $L = 8$ this situation occurs rarely, cf. Schulze and Stauffer 2006). Simulations using 64 bits, of which 19 bits are reserved for families and the remaining 45 bits for languages were also made (allowing for $2^19 = 524,288$ possible `families' and $2^45 = 35,184,372,088,832$ possible `languages'). The results were qualitatively similar.

Differentiation is simulated by setting the probability of the change in a bit to 0.0001 per iteration. An iteration is equivalent to a certain, average time step. After some time steps, a bit in either the family sub-string or the language bit-string will change, meaning the creation of a new entity at one of these levels. Given that there are fewer bits in the family bit-string, there is a smaller probability of a change in this part of the string per iteration, and there will therefore be a slower dynamics of families than of languages. In practice, with probability $0.0001 L$ at each iteration, one of the $L$ bits is selected randomly and then reverted, i.e. changed from 0 to 1 or from 1 to 0. In this process, analogously to biological mutations, all bit positions are equivalent and neither 0 nor 1 is in any way preferred.
 	
We neglect here for simplicity the diffusion of features from one language to the other used in other simulations involving this model. We assume a shift from small to large populations stipulating that at each iteration with probability $(1-x)^2$ or $(1-x^2)$ each individual gives up his/her old language and instead selects the language of one randomly selected individual of the whole population. Individuals get one child per iteration, and everybody dies with a Verhulst probability proportional to the current population size, something which takes into account factors such as limited food and space. We usually start with a population corresponding roughly to the equilibrium size determined by these Verhulst deaths, where everybody starts with a randomly selected language. After some time, one language may dominate and be spoken by more than 80 percent of the population. Stauffer et al. (2006a) list a complete Fortran program. The histograms of the number of languages spoken by a given number of people are smoothened by random multiplicative noise as in Stauffer et al. (2006b), which may correspond to external perturbations caused by migrations of individuals, intermarriage, changing political circumstances, and other non-systematic factors.

\subsection{The flat variant}

This model is in all but one major respect similar to the hierarchical model. The difference is that taxonomic levels are achieved not by partitioning the string, but by stipulating that two languages which differ in more than one bit belong to different families. The size of each language family (i.e., the numbers of languages in each) is then measured by the number of languages that differ by just one bit from one reference language. We sum over all reference languages, and also over many samples, to get out final statistics. The definition allows one language to belong to different families, just as one person can belong to different friendship groups. Instead, one would get a clear separation into different families without such overlaps if we demand all languages within one family to be separated directly or indirectly by not more than one bit flip. But since we can move from each bit-string of 64 bits to every other possible bit-string through at most 64 such changes of single bits, this definition would mean that all possible languages form one huge family, which is not what we want. (Analogously, on a square lattice we can define a neighbourhood as the set of four nearest neighbours of a given site; then every lattice site belongs to several neighbourhoods. Alternatively, a cluster can be defined as the set of all sites connected directly or indirectly with a given site; then the whole lattice forms one large cluster. Neither definition leads to what we would like to have, which is non-overlapping clusters, corresponding to non-overlapping language families. A further disadvantage of the model is that its equilibrium is either dominance of one language spoken by most people, or fragmentation into numerous languages of about equal size; thus for dominance there is not much to analyze and for fragmentation nearly all languages could form one cluster, meaning that these more sophisticated definitions might not work better in equilibrium.)

\section{Results}

\subsection{Results for the application of the hierarchical model}

The major results are shown in figs. 1-2 and 4-5. The interest of these are the shapes of the various curves, not the absolute numbers corresponding to each point. The mismatch between large numbers of families and small sizes of languages as compared to the real-world situation is due to the summation over iterations and could be normalized, but this would only serve presentational purposes.

In fig. 1 it is shown how size histograms of families strongly depend on the temporal factor. At the initial stage of the simulation ($t = 1$) we see something close to a normal distribution (the rightmost curve in the diagram). At $t = 10$ the distribution forms a parabola (curve connecting x's). This distribution is close to what the present-day \textit{language} size distribution looks like (see fig. 6). At $t = 60$ (stars) a curve resembling the present-day distribution of \textit{language family} sizes (fig. 3) is obtained, but it has a large hump on the right region of the curve. The real-life distribution also has a hump, but it is much smaller. At 300 iterations (squares) there is a discontinuous distribution with a number of small families and a leap up to a number of larger ones, which form a narrow normal distribution. Fig. 2a focuses on the range $20 \le t \le 150$, where the distribution most closely resembles the present-day one, and varies the population size $N$ to see the dependency on the graph on that variable. It appears that there is not much influence of $N$, provided $t$ is increased with increasing $N$. Moreover, fig. 2a suggests that the closest approximation to the present-day distribution is found around $10^2$ iterations. Statistically solid results for a long run of the 64-bit model in fig. 2b provide similar results.

We now turn to the results for language sizes. Fig. 4 shows the sizes for the same number of iterations as in fig. 1. Since the simulations start with fragmentation, $t = 1$ represents a situation with many languages spoken by single speakers (single +). At $t = 10$ (x symbols) a curve roughly like a parabola and already strongly reminiscent of the present-day situation (fig. 6) has begun to form. At $t = 60$ (stars) this distribution is beginning to disrupt, as evidenced by the right tail. This situation further develops into one with many large languages and many small ones, with a large gap for language sizes in between, as shown by the curve for $t = 300$ (squares). Again we narrow in on the range, $20 \le t \le 60$, where the best approximation of the present situation (fig. 6) is found and vary the population size (fig. 5). For $t = 40$ and $N = 50,000$ the distribution closely approximates the present-day one.  

By comparing the curves for $t = 40$ in figs. 2a and 5 an interesting observation is obtained: at identical time steps the curve for language family sizes may approximate a power-law while the curve for language sizes does not, but rather something close to a parabola, as in real life. Wichmann (2005: 128) hypothesized that both curves should approximate a power-law, but the simulations rather suggest that this is only the case for language family sizes, at least given the model and the setting of parameters assumed here.
 
The overall result, then, suggests that neither the present-day distribution of language family sizes nor that of language sizes are unexpected and that both may have been obtained for a long time and may continue to be obtained. Eventually a dominance of just one large language accompanied by other slightly different languages is possible, but this situation has not yet set in.

\subsection{Results for the application of the flat model}

For investigating the distance among languages the `flat' model is most useful because the distance among two languages belonging to two different families in the hierarchical model cannot easily be measured. (The hierarchical bit-strings representing languages in any two languages belong to two different families are not comparable since the positions no longer mean the same when one moves up one taxonomic level.) Thus we measured differences in simulations implementing the non-hierarchical model, i.e. the standard model of Schulze and Stauffer (2005, 2006) where all bits are equivalent. As in most of our previous studies, only short bit-strings of 8 or 16 bits were used, and the random multiplicative noise was omitted; for these studies we waited until a stationary state after about $10^3$ iterations was established.

The distance measure used is the so-called `Hamming distances', also investigated by Te\c sileanu and Meyer-Ortmanns (2006). The Hamming distance between two bit-strings is the number of bits which are different in a position-by-position comparison of the two strings. For example, the Hamming distance between 01001101 and 11000011 is four.

As explained above, we define a language family in this model as a set of languages differing from a given reference language by not more then $k$ bits, in this case setting $k$ to one bit. The results of the simulations of bit-strings of lengths 8 and 16 are shown in fig. 7; as in fig. 2 above, the simulations represent states of non-equilibrium, i.e., they were stopped at some intermediate time and not let run until the distribution no longer changed apart from random fluctuations. These results are not very different from those shown in fig. 2 for the 64 bits string in the hierarchical model.

\subsection{More on Hamming distances}

The above results were obtained by stopping the simulations at a suitable time such that the results are closest to reality. In this section we report on the equilibrium properties for longer times where the distributions no longer
change appreciably and where we will have either dominance of one language or fragmentation of the whole population into many different languages.

Fig. 8 nicely shows the phase transition between dominance at low and fragmentation at high mutation rate $p$ per bit-string when we vary the mutation rate instead of fixing it to only 0.0001 mutations per bit and per iteration. For dominance, nearly everybody speaks one language, and most of the others speak a language differing in only one bit from this dominating language. Fragmentation happens for larger mutation rates; then all possible languages are represented about equally. We see in fig. 8 that dominance is characterized by a small average Hamming distance while for fragmentation the average Hamming distance is about 1/2 (here it is normalized by the length of the bit-string such that two random bit-strings have on average a distance 1/2.) This effect is already seen if one looks only at the two largest languages in the population, as done by Te\c sileanu and Meyer-Ortmanns (2006).

For fragmentation, the distribution of Hamming distances between two pairs of speakers is roughly Gaussian (normal), shown by a parabola in the semi-logarithmic plot (stars in fig. 8). In the case of dominance, as observed for two lower mutation rates $p$ in fig. 9, the most probable Hamming distance is zero, and for higher distances the probability to observe them decays very rapidly. 

In these simulations we started with one language only and used the probability $1-x^2$ for the shift from small to large languages. We got qualitatively similar results when we started from a population fragmented into many languages, except that then the probability of a shift was set to $(1-x)^2$, to allow for a possible transition from fragmentation to dominance.

\section{Conclusion}

The primary aim of our simulations was to capture, within one and the same model, how two different empirically observed distributions might arise, i.e. a roughly log-normal distribution of language sizes and an approximate power-law for the family sizes. With reasonable lengths of bit-strings, populations and observation times we could, indeed, find the two different behaviours in the same simulation. This suggests, contrary to the hypothesis of Wichmann (2005), that the present-day distribution of language family sizes in combination with that of language sizes may not be unexpected.

In terms of simulation techniques the major contribution of the present paper has been the introduction of new models into the area of linguistic taxonomic dynamics, an area which, to our knowledge, has not previously been investigated by means of computer simulations. The best results were obtained in implementations of the hierarchical bitstring, a model which also has the advantage of being versatile and easy to implement.

The investigations, however, also revealed some problems with the model. If for a fixed length of the bit-strings the population size N goes to infinity, then in the parameter region of fragmentation all possible languages will be spoken, and all possible families will exist, making taxonomy a mathematical triviality without connection to reality. Thus simulations of large but finite populations, as presented here, may be better than mathematically exact solutions for infinite populations. Moreover, we did find an effective power-law for the family size distribution, but that distribution decayed much faster with increasing number of languages than the real distribution, shown fig. 3. Thus future research should aim at also applying and testing other models, such as that of de Oliveira et al. (2006a,b), to problems of linguistic taxonomic dynamic.

\bigskip
\bigskip
\noindent
{\bf \Large References}

\bigskip
\noindent
Abrams, Daniel and Steven H. Strogatz. 2003. Modelling the dynamics of language death. \textit{Nature} 424: 900.

\medskip
\noindent
Axelrod, Robert. 1997. The dissimination of culture: a model with local convergence and global polarization. \textit{The Journal of Conflict Resolution} 41: 203-226.

\medskip
\noindent
\textit{Ethnologue: Languages of the World} (14th edn. edited by Grimes, Barbara F.  2000, 15th edition edited by Raymond, G. Gordon 2005). Dallas, TX: Summer Institute of Linguistics.

\medskip
\noindent
Gomes, Marcelo A. F., G. L. Vasconcelos, I. J. Tsang, and Ing Ren Tsang. 1999. Scaling relations for diversity of languages. \textit{Physica A} 271: 489-495.

\medskip
\noindent
Nettle, Daniel. 1999a. Linguistic diversity of the Americas can be reconciled with a recent colonization. \textit{Proceedings of the National Academy of Sciences of the U.S.A.} 96: 3325-3329.

\medskip
\noindent
Nettle, Daniel. 1999b. Using social impact theory to simulate language change. \textit{Lingua} 108: 95-117.

\medskip
\noindent
Newman, Mark E. J. 2005. Power laws, Pareto distributions and Zipf's law. \textit{Contemporary Physics} 46: 323-351.

\medskip
\noindent
Novotny, Vojtech and Pavel Drozd. 2000. The size distribution of conspecific populations. \textit{Proceedings of the Royal Society of London} B267: 947-952.

\medskip
\noindent
Oliveira, Viviane M. de, Marcelo A. F. Gomes, and Ing Ren Tsang. 2006a. Theoretical model for the evolution of the linguistic diversity. \textit{Physica A} 361: 361-370.

\medskip
\noindent
Oliveira, Viviane M. de, Paulo R. A. Campos, Marcelo A. F. Gomes, and Ing Ren Tsang. 2006b. Bounded fitness landscapes and the evolution of the linguistic diversity, e-print physics 0510249 for \textit{Physica A}.

\medskip
\noindent
Patriarca, Marco and Teemu Lepp\"anen. 2004. Modeling language competition. \textit{Physica A} 338: 296-299.

\medskip
\noindent
Schulze, Christian and Dietrich Stauffer. 2005. Monte Carlo simulation of the rise and fall of languages. \textit{International Journal of Modern Physics C} 16: 781-787.

\medskip
\noindent
Schulze, Christian and Dietrich Stauffer. 2006. Computer simulation of language competition by physicists. In: Chakrabarti, B. K., A. Chakraborti and A. Chatterjee (eds.), \textit{Econophysics and Sociophysics: Trends and Perspectives}. Weinheim: WILEY-VCH Verlag; and: Recent developments in computer simulations of language competition, Computing in Science and Engineering  8 (May/June) 86-93.

\medskip
\noindent
Stauffer, Dietrich, Suzana Moss de Oliveira, Paulo Murilo C. de Oliveira, Jorge S. S\'a Martins. 2006a. \textit {Biology, Sociology, Geology by Computational Physicists}. Amsterdam: Elsevier.

\medskip
\noindent
Stauffer, Dietrich, Christian Schulze, F. Welington S. Lima, Søren Wichmann, and Sorin Solomon. 2006b. Non-equilibrium and irreversible simulation of competition among languages. \textit{Physica A}. (In press).

\medskip
\noindent
Sutherland, William J. 2003. Parallel extinction risk and global distribution of languages and species. \textit{Nature} 423: 276-279.

\medskip
\noindent
Tes\c sileanu, Tiberiu and Hildegard Meyer-Ortmanns. 2006. Competition and languages and their Hamming distance. arXiv:physics/0508229, \textit{International Journal of Modern Physics C} 17: 259-278.

\medskip
\noindent
Wang, William S. Y. and James W. Minett. 2005. The invasion of language: emergence, change and death. \textit{Trends in Ecology and Evolution} 20.5: 263-296.

\medskip
\noindent
Wichmann, S{\o}ren. 2005. On the power-law distribution of language family sizes. \textit{Journal of Linguistics} 41: 117-131.

\begin{figure}[hbt]
\begin{center}
\includegraphics[angle=-90,scale=0.5]{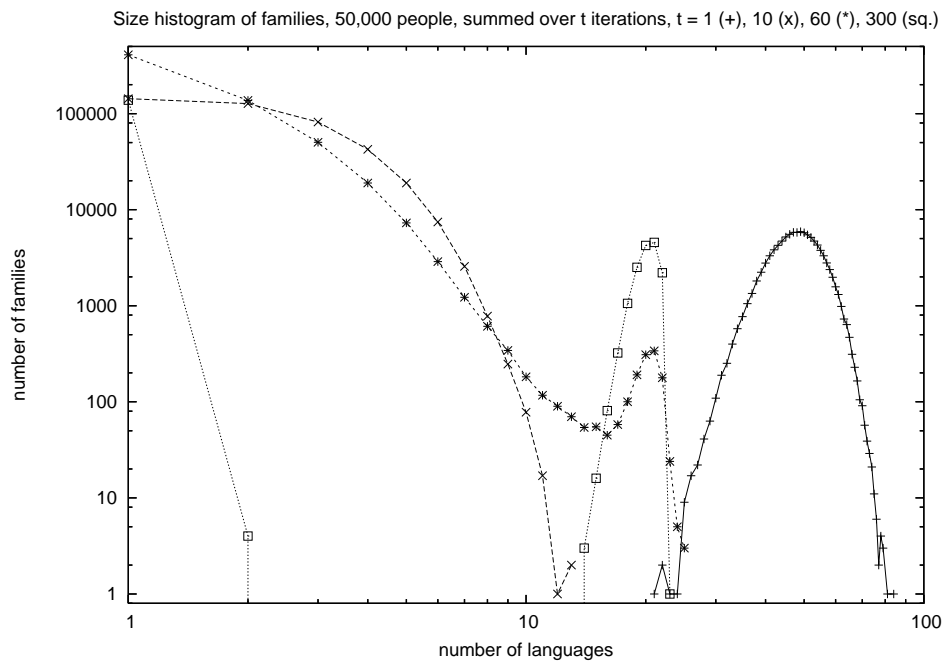}
\end{center}
\caption{Time dependence of the distribution of family sizes, summed over 100 samples at $L = 32$. For long times a narrow peak develops, shown by squares.
}
\end{figure}

\begin{figure}[hbt]
\begin{center}
\includegraphics[angle=-90,scale=0.45]{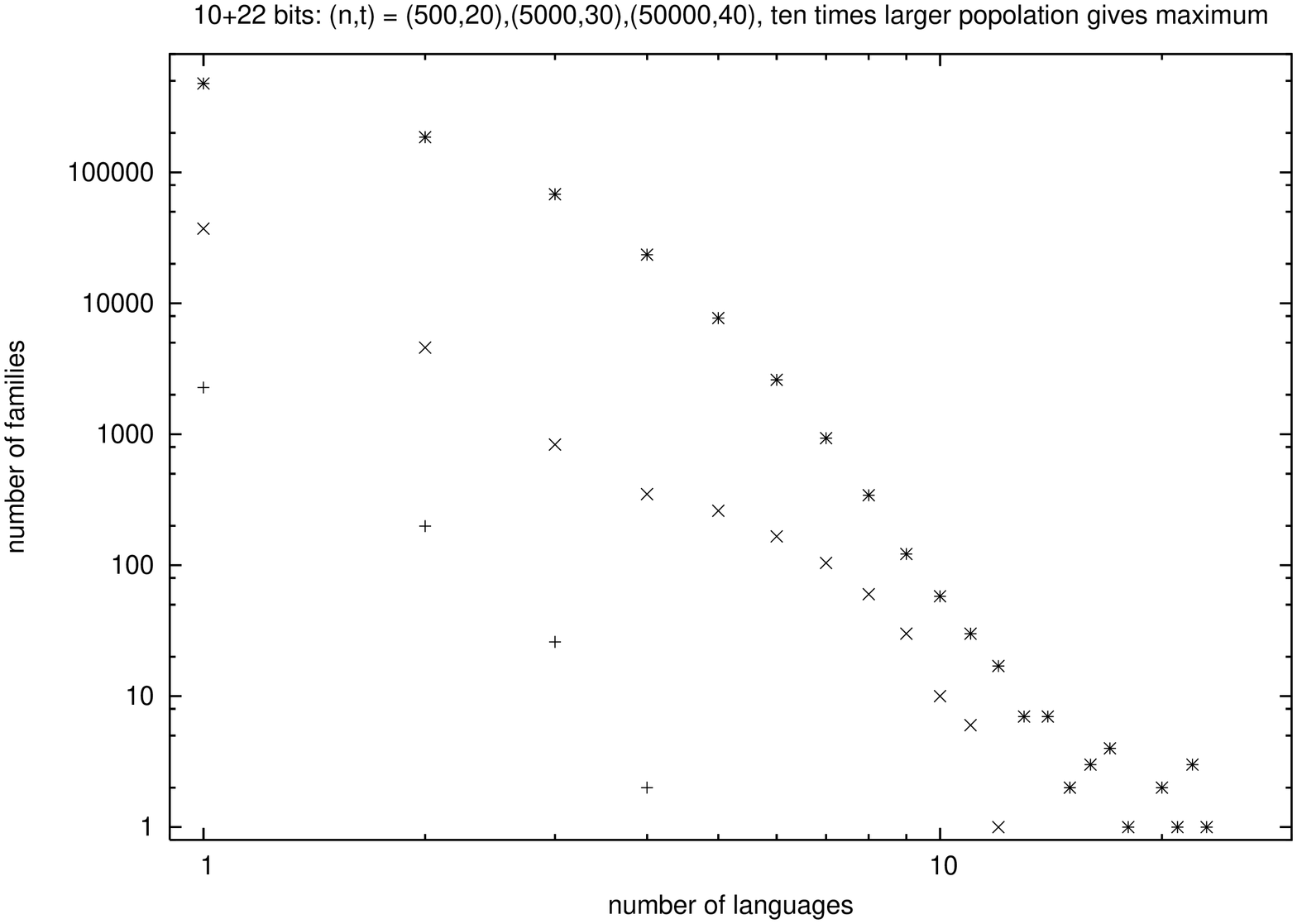}
\includegraphics[angle=-90,scale=0.45]{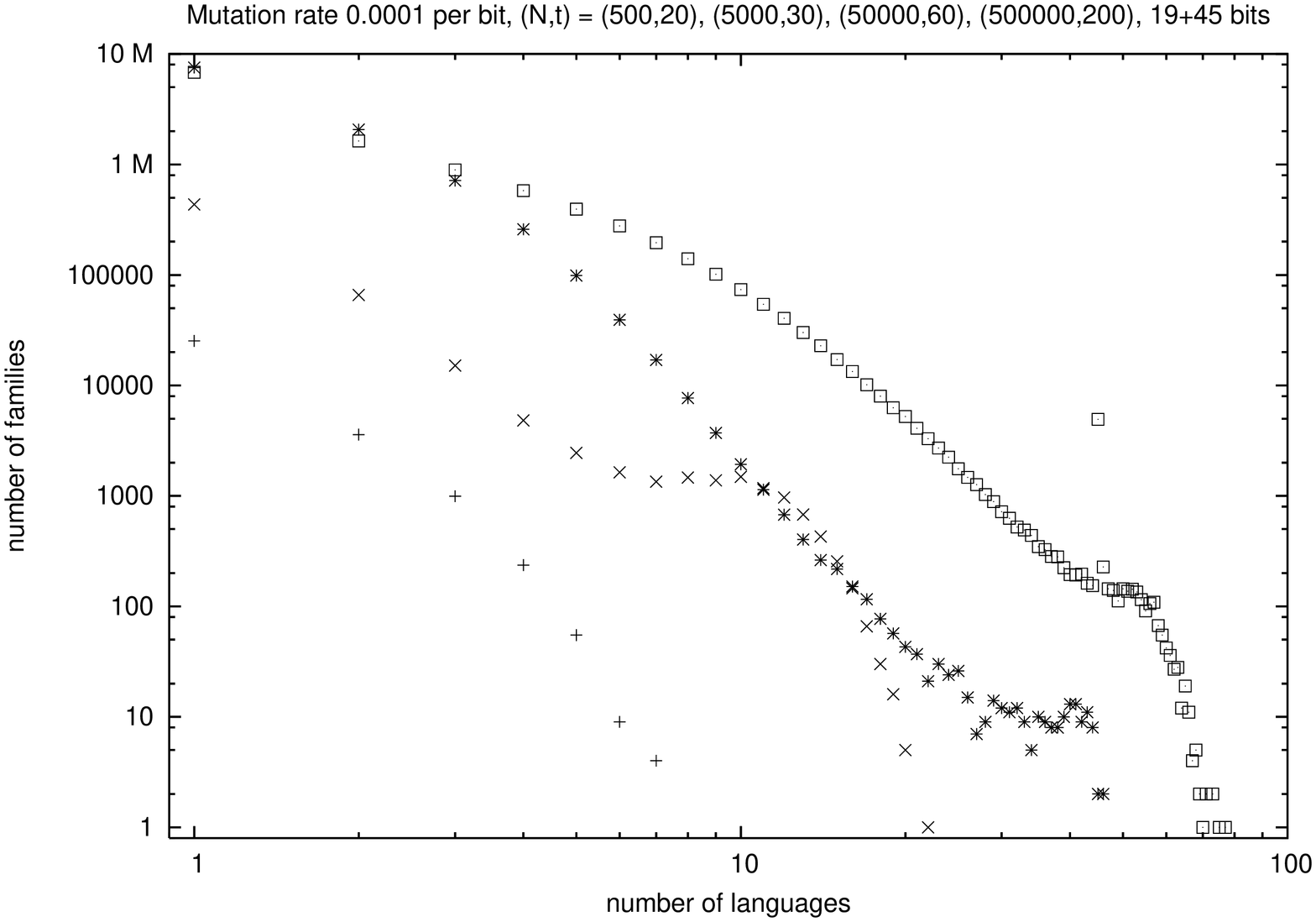}
\end{center}
\caption{Population-size dependence of the distribution of family sizes, summed over 100 samples, at $L = 32$ (top) and 64 (bottom) and for suitably selected intermediate times. These results are roughly independent of the population.
}
\end{figure}

\begin{figure}[hbt]
\begin{center}
\includegraphics[angle=-90,scale=0.34]{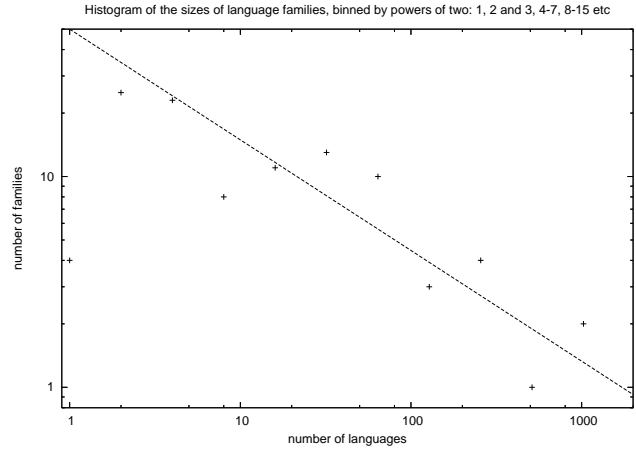}
\end{center}
\caption{Empirical distribution of family sizes, from {\textit Ethnologue};
see also Wichmann (2005: fig. 2).
}
\end{figure}

\begin{figure}[hbt]
\begin{center}
\includegraphics[angle=-90,scale=0.34]{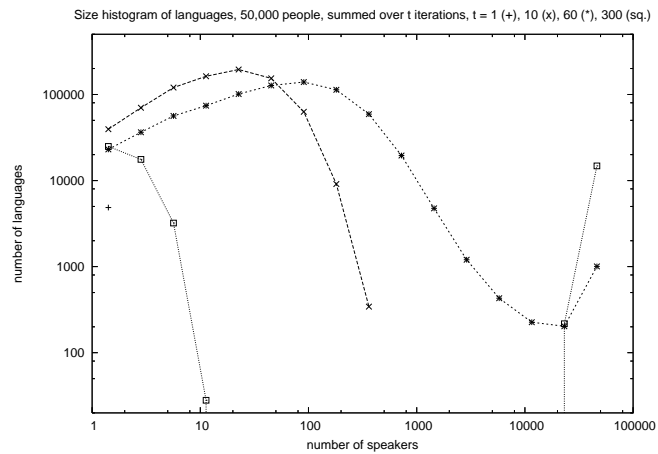}
\end{center}
\caption{Time dependence of the distribution of language sizes, summed over 100
samples at $L = 32$. For long times dominance of one language develops, 
leading to an isolated peak at language sizes slightly below the total
population size. Only intermediate times give the desired roughly parabolic
shape. The same simulations were used for these language sizes as for the
family sizes in fig. 1.
}
\end{figure}

\begin{figure}[hbt]
\begin{center}
\includegraphics[angle=-90,scale=0.34]{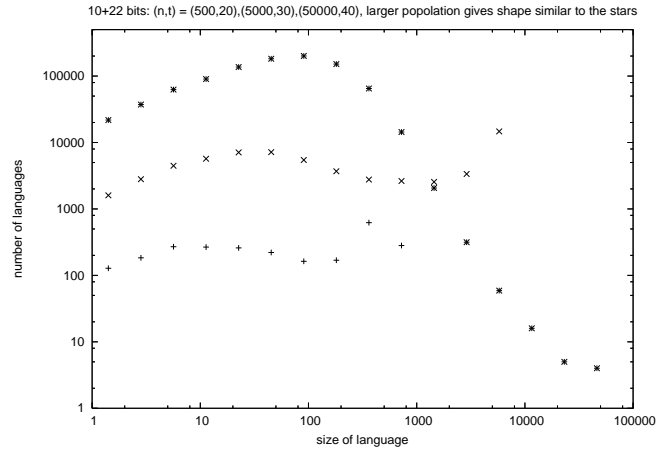}
\end{center}
\caption{Population-size dependence of the distribution of language sizes.
Same simulations as in fig. 2a for family sizes. 
}
\end{figure}

\begin{figure}[hbt]
\begin{center}
\includegraphics[angle=-90,scale=0.34]{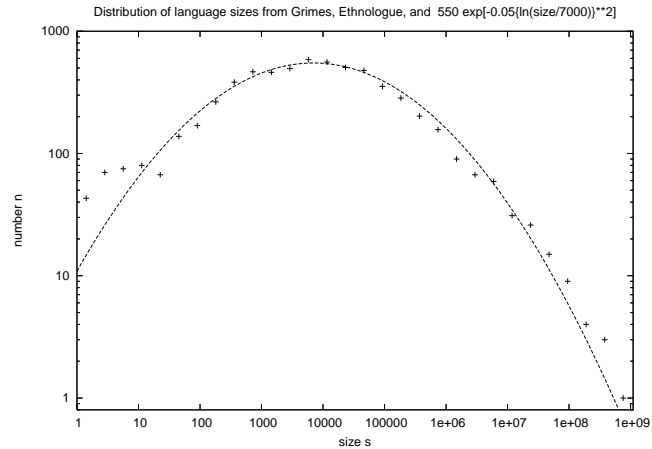}
\end{center}
\caption{Empirical distribution of language sizes, from \textit{Ethnologue};
see also Sutherland (2003: fig. 1), Wichmann (2005: fig. 6).
}
\end{figure}

\begin{figure}[hbt]
\begin{center}
\includegraphics[angle=-90,scale=0.34]{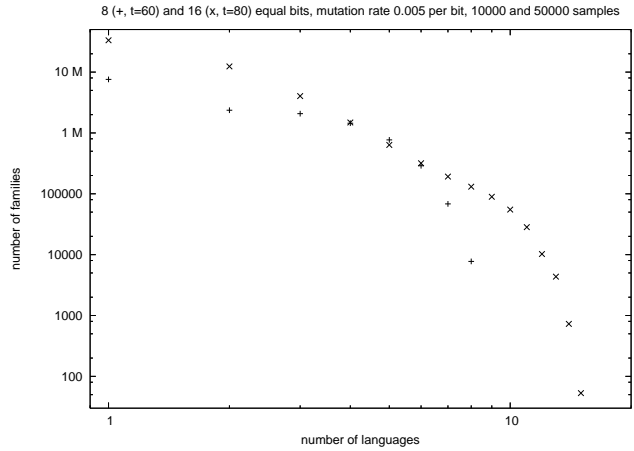}
\end{center}
\caption{Flat version: distribution of family sizes for 8 and 16 bits.
}
\end{figure}

\begin{figure}[hbt]
\begin{center}
\includegraphics[angle=-90,scale=0.34]{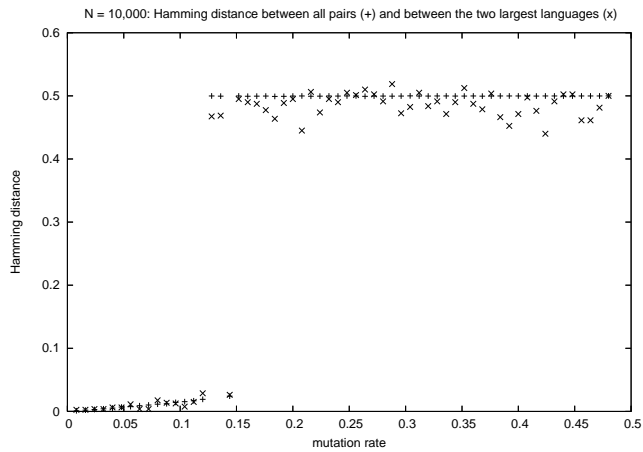}
\end{center}
\caption{The average normalized Hamming distance ifor $L = 8$ jumps
from low values (dominance) to nearly 1/2 (fragmentation) when the mutation
rate increases.
}
\end{figure}

\begin{figure}[hbt]
\begin{center}
\includegraphics[angle=-90,scale=0.5]{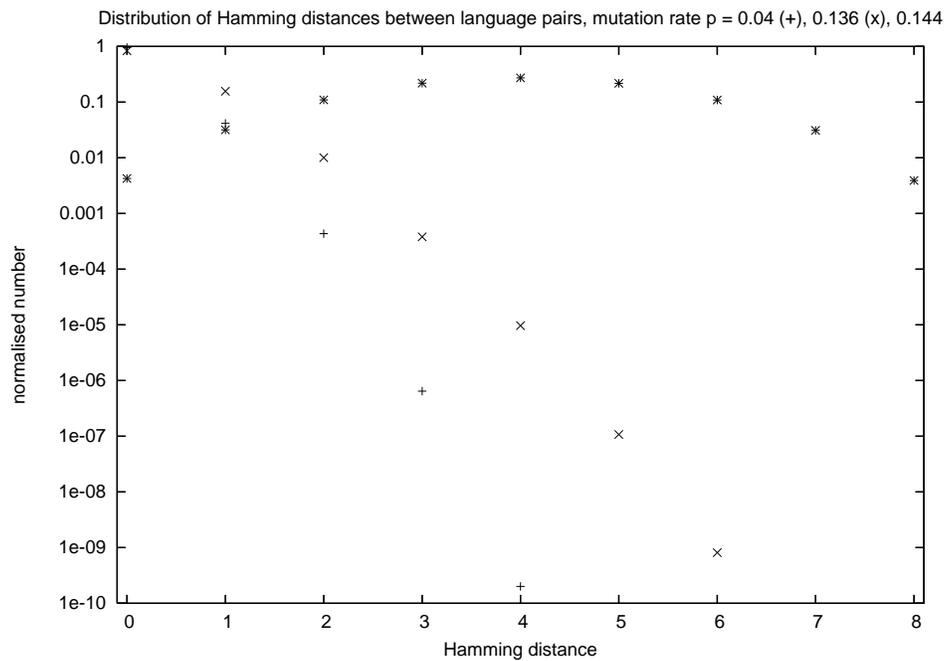}
\end{center}
\caption{Flat version: distribution of $k$ values, where $k$ is the Hamming
distance between an arbitrary pair of existing languages, at $L = 8$. The 
parabolic maximum corresponds to fragmentation at a high mutation rate, 
the two rapidly decaying curves to dominance at lower mutation rates.
}
\end{figure}

\end{document}